\documentclass[aps, pra, reprint, longbibliography, superscriptaddress, amsmath, amssymb, floatfix, showkeys]{revtex4-2}

\usepackage{graphicx}
\usepackage{amsmath,amssymb,mathtools,bm}
\usepackage[colorlinks=true,linkcolor=blue,citecolor=blue,urlcolor=blue]{hyperref}

\DeclarePairedDelimiter\ket{\lvert}{\rangle}
\DeclarePairedDelimiterX\braket[2]{\langle}{\rangle}{#1\,\delimsize\vert\,\mathopen{}#2}

\begin{document}

\title{Krylov complexity of a tilted extended Bose-Hubbard chain with Rydberg-dressed interactions}

\author{Yifan Chen}
\affiliation{School of Physics and Astronomy, University of Nottingham, Nottingham NG7 2RD, United Kingdom}
\affiliation{Taishan College, Shandong University, Jinan 250100, China}

%\author{Yinuo Li}
%\affiliation{School of Physics and Astronomy, University of Nottingham, Nottingham NG7 2RD, United Kingdom}
%\affiliation{Taishan College, Shandong University, Jinan 250100, China}

\author{Tianyi Yan}
\affiliation{School of Physics and Astronomy, University of Nottingham, Nottingham NG7 2RD, United Kingdom}
\affiliation{Centre for the Mathematics and Theoretical Physics of Quantum Non-equilibrium Systems, University of Nottingham, Nottingham NG7 2RD, United Kingdom}

\author{Lu Qin}
\affiliation{School of Physics, Henan Normal University, Xinxiang 453007, China}
\affiliation{School of Physics and Astronomy, University of Nottingham, Nottingham NG7 2RD, United Kingdom}
\affiliation{Centre for the Mathematics and Theoretical Physics of Quantum Non-equilibrium Systems, University of Nottingham, Nottingham NG7 2RD, United Kingdom}

\author{Weibin Li}
\affiliation{School of Physics and Astronomy, University of Nottingham, Nottingham NG7 2RD, United Kingdom}
\affiliation{Centre for the Mathematics and Theoretical Physics of Quantum Non-equilibrium Systems, University of Nottingham, Nottingham NG7 2RD, United Kingdom}

\begin{abstract}
We investigate Krylov state complexity in a tilted extended Bose-Hubbard chain in which both the on-site interaction $U$ and nearest-neighbor interaction $V$ are present. The tilted extended Bose-Hubbard model can be realized with Rydberg-dressed interactions in optical lattices. Using exact diagonalization 
and Lanczos recursion from three physically motivated Fock states, we compute
the time-dependent complexity, its long-time saturation value, and the fluctuations of the Lanczos coefficients for an open chain at unit filling. Crucially, we identify a pronounced quasi-chaotic diagonal
regime near $U\simeq V$ in which the Krylov complexity saturation exhibits a sharp ridge while the spectral level-spacing statistics remain partially chaotic. This dissociation
arises from the competition between the on-site and nearest-neighbor interaction, which generates a dense manifold of near-resonant many-body configurations that is
efficiently coupled by the state-dependent dynamics but retains residual spectral structure.
Our results show that Krylov state complexity provides complementary, state-sensitive information about many-body chaos that is not captured by the spectral diagnostics
alone, establishing Krylov complexity as a useful diagnostic for studying thermalization and information scrambling in experimentally accessible quantum simulators.
\end{abstract}

\maketitle

\section{Introduction}

Ultracold atoms in optical lattices provide a highly controllable platform for studying
strongly correlated lattice models, nonequilibrium many-body dynamics, and the crossover
between regular and chaotic behavior~\cite{Bloch2008}. The Bose--Hubbard model is the
paradigmatic example that describes the competition between tunneling and on-site
interactions that supports the superfluid--Mott-insulator transition~\cite{Fisher1989,Jaksch1998,Greiner2002}. Extensions of this model arise
naturally when the interaction range exceeds a single lattice site, giving rise to extended Bose-Hubbard model. The extended Bose-Hubbard model supports density-wave
and supersolid states, richer phase diagrams, and more intricate excitation
spectra~\cite{Dutta2015,Mishra2009}. Such nonlocal interactions can be engineered with
dipolar particles~\cite{Lahaye2009,Baier2016} or with Rydberg dressing~\cite{Henkel2010,Pupillo2010,johnsonInteractionsRydbergdressedAtoms2010}. The latter is largely benefited from the strong and long-range interaction of atoms in electronically high-lying Rydberg states~\cite{Saffman2010,Browaeys2020,shaoRydbergSuperatomsArtificial2024a}, offering a particularly flexible route to tunable long-range
interactions in quantum simulators~\cite{liProbingInteractionRydbergdressed2012,PhysRevLett.108.265301,PhysRevA.86.013619,PhysRevA.94.053422,PhysRevLett.114.243002,Zeiher2016,PhysRevA.98.063635,schaussQuantumSimulationTransverse2018,liSupersolidityLatticeBosons2018,zhouQuenchDynamicsRydbergdressed2020,mccormackNonlinearDynamicsRydbergdressed2020,GuardadoSanchez2021,PhysRevA.102.063307,PhysRevA.104.053304,PhysRevA.102.043308,wuConciseReviewRydberg2021,PhysRevA.104.L061302,wuFastSpinSqueezing2022,PhysRevA.108.053107,PhysRevResearch.2.023290,cardarelliAccessingTopologicalMott2023,Chen:23,wangSpinOrbitCoupled2024,PhysRevB.109.054519,PhysRevA.111.043318,PhysRevResearch.6.023151}. Rydberg dressed interactions have been experimentally demonstrated~\cite{hollerithRealizingDistanceSelectiveInteractions2022,hinesSpinSqueezingRydberg2023}, and the extended Bose-Hubbard model with the Rydberg dressed interaction has been experimentally realized~\cite{weckesserRealizationRydbergdressedExtended2025}. The rapid experimental developments motivate the exploration of novel ground state phases and many-body dynamics driven by the long-range Rydberg dressed interaction.

Quantum chaos in interacting many-body systems is most commonly characterized through spectral probes. Integrable spectra are expected to show Poisson-distributed level spacings~\cite{Berry1977}, whereas the Bohigas-Giannoni-Schmit conjecture links level repulsion to random-matrix universality in quantum systems whose classical analogs are chaotic~\cite{Bohigas1984}. In lattice many-body settings, adjacent-gap statistics and related measures provide robust indicators of the integrable-to-chaotic crossover~\cite{Oganesyan2007,Atas2013,Santos2010}. Dynamically the system typically thermalizes as predicted by the eigenstate thermalization hypothesis (ETH)~\cite{Deutsch1991,Srednicki1994,Rigol2008,DAlessio2016}. The Bose-Hubbard model itself has long been recognized as a fertile setting for many-body quantum chaos~\cite{Kolovsky2004,Hiller2006}. Nevertheless, spectral and eigenstate diagnostics have known limitations in finite systems. For example, they depend on symmetry resolution, can fluctuate strongly when the Hilbert space is modest, and provide no direct prediction about how rapidly quantum information is scrambled.

A complementary, dynamical route to quantum chaos runs through the spreading of quantum information under time evolution. Out-of-time-ordered correlators provide a direct probe of information scrambling and a fundamental bound on the quantum Lyapunov exponent~\cite{Maldacena2016}. Krylov complexity offers a closely related but distinct framework. In the operator formulation~\cite{Parker2019,Rabinovici2021}, Lanczos recursion generates a basis adapted to the Heisenberg-evolved operator.  The resulting complexity captures how far the operator has spread from its initial form and is conjectured to grow at most linearly in time for generic chaotic systems, with the early exponential growth rate bounded by the Lyapunov exponent. Bounds of the Lanczos growth are identified~\cite{Parker2019,Avdoshkin2020}.  In the state  formulation~\cite{Balasubramanian2022}, the same recursion is applied to evolution of the quantum state, yielding a measure of how broadly the initial wavefunction has spread over the Krylov basis. Complexities play important roles in the dynamics of  different models~\cite{Dymarsky2021,Rabinovici2022,Heveling2022,BallarTrigueros2022,hashimoto2023}

Krylov state complexity has been shown to capture the integrable-to-chaotic crossover and to respond sensitively to localization~\cite{Scialchi2024,Baggioli2025}. It shows characteristic early quadratic growth, a linear ramp to a sharp peak, and then relaxation to a plateau in ergodic systems. The role of the initial state in determining Krylov saturation has been examined~\cite{SreeramPG2025}. The relation of Krylov complexity with quantum chaos~\cite{Balasubramanian2025} and interaction scrambling~\cite{Chen2025} has been studied too. Krylov operator complexity has been computed in three-site Bose-Hubbard models~\cite{Bhattacharyya2023}, showing clear signatures of chaos in the strong-interaction regime.  However, Krylov state complexity in extended Bose-Hubbard chains with competing interaction ranges has received limited attention. In viewing the recent experiment achievement~\cite{weckesserRealizationRydbergdressedExtended2025}, this motivates us to study the Krylov complexity in the extended Bose-Hubbard model.

In this work we analyze Krylov state complexity in a tilted one-dimensional extended Bose-Hubbard chain at unit filling. The weak tilt removes reflection symmetry, so the level statistics do not require a separate parity resolution~\cite{kollathStatisticalPropertiesSpectrum2010}. We identify three main findings: (i) enhanced Krylov spreading at intermediate interactions when the hopping $J$ is comparable to the nearest-neighbor interaction $V$. It distinguishes chaotic from integrable regimes, consistent with recent theoretical predictions for bosonic systems~\cite{Bhattacharyya2023,Scialchi2024}; (ii) a pronounced diagonal ridge near $U\simeq V$ in the saturation of the Krylov complexity that is not accompanied by an equally sharp spectral signature, indicating strong state-dependent information scrambling due to the competing interaction, and (iii) demonstration that the Krylov state complexity provides complementary information to level-spacing statistics and eigenstate delocalization, making it a valuable diagnostic for experimentally accessible quantum simulators~\cite{weckesserRealizationRydbergdressedExtended2025}. 

%The diagonal quasi-chaotic regime represents the central discovery.  We identify it as a finite-size signature of many-body dynamics where competing interactions generate efficient Krylov-space transport.

The paper is organized as follows. Section~\ref{sec:model} introduces the model, the initial states, and the Krylov and spectral observables. Section~\ref{sec:results} presents the interaction-plane overview and the figure-by-figure discussion of the main parameter cuts, followed by an interpretation of the diagonal anomaly. In Section~\ref{sec:discussion}, we discuss the connection of the diagonal ridge with the parameters, and the ground-state properties. Section~\ref{sec:conclusion} summarizes the main conclusions and outlines future directions.

\section{Model and complexity diagnostics}
\label{sec:model}

\subsection{Tilted extended Bose-Hubbard chain}

We consider a one-dimensional extended Bose-Hubbard model describing $N$ bosonic atoms confined to a chain of $L$ sites with open boundary conditions, given by 
 Hamiltonian ($\hbar = 1$)
\begin{align}
\hat{H}={}&-J\sum_{i=1}^{L-1}\left(\hat{a}_{i}^{\dagger}\hat{a}_{i+1}+\hat{a}_{i+1}^{\dagger}\hat{a}_{i}\right)
+\frac{U}{2}\sum_{i=1}^{L}\hat{n}_{i}(\hat{n}_{i}-1)\notag\\
&+V\sum_{i=1}^{L-1}\hat{n}_{i}\hat{n}_{i+1}
+F\sum_{i=1}^{L}\left(\left\lfloor\frac{L}{2}\right\rfloor+1-i\right)\hat{n}_{i},
\label{eq:H}
\end{align}
where 
$\hat{a}_{i}^{\dagger}$ and $\hat{a}_{i}$ are bosonic  creation and annihilation operators at site $i$, and $\hat{n}_{i}=\hat{a}_{i}^{\dagger}\hat{a}_{i}$ is the number operator. 
$J$ is the nearest-neighbor hopping rate, and $\left\lfloor\cdot \right\rfloor$ denotes the floor function.
The on-site interaction $U=4\pi\hbar^2 a_s/m$ (i.e., $U=4\pi a_s/m$ for
$\hbar=1$) is set by the $s$-wave scattering length $a_s$ and the atomic mass $m$, and can be
tuned via a Feshbach resonance~\cite{pethick2002bose,stringari2016bose,Makotyn2014universal}. The
nearest-neighbor interaction $V$ originates from the soft-core potential generated by Rydberg
dressing, $V={C_6}/{(d^6+R_c^6)}$,
%\begin{equation}
%V=\frac{C_6}{d^6+R_c^6},\notag
%\end{equation}
where $C_6$, $d$, and $R_c$ are the effective dispersion coefficient, the lattice constant,
and the soft-core radius, respectively~\cite{Henkel2010,Pupillo2010,liProbingInteractionRydbergdressed2012}. The soft-core interaction is strong when $R_c\sim d$, and decays rapidly with larger distances. We work in this regime such that even longer-range interaction can be neglected. Finally, $F$ is the energy bias between neighboring sites, i.e., the linear tilt. We choose $F=0.1J$, which is sufficiently weak that it does not qualitatively induce Wannier–Stark localization~\cite{Hartmann_2004}. The weak tilt explicitly breaks spatial reflection symmetry and thereby removes symmetry-related degeneracies and accidental level crossings in the finite  chain setting. In the remainder of the article, the hopping amplitude $J$ will be used to set the energy scale.

For unit filling and in the fixed-$N$ sector, the Hilbert-space dimension is $\mathcal{D}=[{(N+L-1)!}]/[{N!\,(L-1)!}]$.
%\begin{equation}
%\mathcal{D}=\frac{(N+L-1)!}{N!\,(L-1)!}.\notag
%\end{equation}
All spectral diagnostics are evaluated within this fixed
particle-number sector. We will focus on $N=L=6$ which gives $\mathcal{D}=462$. In this way, the resulting Hilbert space is large enough to exhibit
nontrivial Krylov complexity, while
remaining tractable for full exact diagonalization. For level-spacing statistics analysis, we will consider $N=12$ in order to get a better statistics. Previous studies have shown that novel dynamics can be found in even shorter Bose-Hubbard chains in both classical and quantum regimes~\cite{Kolovsky2004,PhysRevA.84.063609,mccormackHyperchaosBoseHubbardChain2021,nakerstChaosThreesiteBoseHubbard2023,yanSignaturesQuantumChaos2023}.

\subsection{Initial states and Krylov complexity}

To study the state dependence of Krylov dynamics, we choose three  Fock states as the initial states,
\begin{align}
\ket{\psi_l}=\ket{600000},~~~
\ket{\psi_h}=\ket{111111},~~~
\ket{\psi_s}=\ket{020202},\notag
\end{align}
where $\ket{n_1 n_2 \cdots n_L}$ denotes the boson occupation number in site $j=1,\cdots, L$.
These are seed states that capture complementary regions of Hilbert space. State $|\psi_l\rangle$ is a maximally localized state with all bosonic atoms on the left site, $|\psi_h\rangle$ is a homogeneous Mott-like product state at unit filling, and $|\psi_s\rangle$ is a period-two staggered density wave state. Because the three states respond differently to on-site and nearest-neighbor interactions, they provide independent dynamical response to different parameter regions of the Hamiltonian.

% These seeds represent a maximally localized configuration with all bosons on the first site, a homogeneous Mott-like product state at unit filling, and a period-two staggered density-wave configuration, respectively.  These three seeds sample complementary regions of Hilbert space and respond differently to on-site versus nearest-neighbor interactions, providing three independent dynamical windows onto the same Hamiltonian.

For each seed  state $j \in \{l,h,s\}$, we take $|K_0\rangle = |\psi_j\rangle$ and generate the Krylov basis via the Lanczos algorithm~\cite{Lanczos1950,viswanath1994recursion}. Setting $|K_{-1}\rangle = 0$ and $b_0 = 0$, the recurrence reads,
\begin{subequations}\label{eq:lanczos}
\begin{align}
&|A_n\rangle = (\hat{H} - a_{n-1})|K_{n-1}\rangle - b_{n-1}|K_{n-2}\rangle,\\
&|K_n\rangle = {|A_n\rangle}/{b_n},
\end{align}
\end{subequations}
for $n \ge 1$. The Lanczos coefficients are
\begin{equation}
\label{eq:lanczos-coeffs}
a_n = \langle K_n | \hat{H} | K_n \rangle, \qquad
b_n = \langle K_{n-1} | \hat{H} | K_n \rangle,
% \langle A_n | A_n \rangle^{1/2},
\end{equation}
and the iteration terminates when $b_n = 0$. In the Krylov basis the Hamiltonian is tridiagonal,
\begin{equation}
\label{eq:tridiag}
\hat{H}|K_n\rangle
=
a_n |K_n\rangle
+ b_n |K_{n-1}\rangle
+ b_{n+1} |K_{n+1}\rangle,\notag
\end{equation}
where $b_K = 0$ for a Krylov space of dimension $K$. If the seed has nonzero overlap with every distinct eigenlevel of $\hat{H}$, then $K = \mathcal{D}$. Otherwise $K$ equals the number of distinct eigenlevels in the spectral associated with the initial state. The numerical results use the resulting
state-dependent Krylov space rather than assuming $K=\mathcal{D}$ a priori.

% Consequently, t
The time evolution
\begin{equation}
\label{eq:time-evolution}
|\psi(t)\rangle
=
e^{-i\hat{H}t}|K_0\rangle
=
\sum_{n=0}^{K-1} \varphi_n(t)|K_n\rangle
\end{equation}
maps onto a one-dimensional tight-binding problem along the Krylov chain, with on-site
potentials $a_n$ and nearest-neighbor hopping amplitudes $b_{n+1}$. The Krylov complexity is
the mean position on this chain~\cite{Parker2019,Balasubramanian2022,zhouDiagnosingQuantumManybody2025},
\begin{equation}
\label{eq:ck}
C_K(t) = \sum_{n=0}^{K-1} n \, |\varphi_n(t)|^2,
\end{equation}
and its long-time average~\cite{Rabinovici2022,PhysRevB.109.224304},
\begin{equation}
\label{eq:cksat}
C_K^{\mathrm{sat}}
=
\lim_{T\to\infty} \frac{1}{T} \int_0^T C_K(t)\,dt
=
\sum_{n=0}^{K-1} n \, Q_{0n},
\end{equation}
is governed by,
\begin{equation}
\label{eq:q0n}
Q_{0n}
=
\sum_{m=1}^{D}
|\langle E_m | K_0 \rangle|^2
\,
|\langle K_n | E_m \rangle|^2,\notag
\end{equation}
where $\{|E_m\rangle\}$ are the exact many-body eigenstates of $\hat{H}$.

At short times, the dynamics is controlled by the energy variance of the initial state,
% ~\cite{hashimoto2023},
\begin{equation}\label{eq:variance}
\sigma^2
=
\langle \psi | \hat{H}^2 | \psi \rangle
-
\langle \psi | \hat{H} | \psi \rangle^2.
\end{equation}
% This quantity measures the magnitude of the erratic behavior of $b_n$.
Because $b_1^2 = \sigma^2$, the complexity grows quadratically,
\begin{equation}
\label{eq:short-time}
C_K(t) = b_1^2 \, t^2 + \mathcal{O}(t^4).
\end{equation}
 Differences in the early-time evolution of $C_K(t)$ therefore reflect the different energy distributions of the three initial states.

\begin{figure*}[ht]
    \centering
    \includegraphics[width=0.97\linewidth]{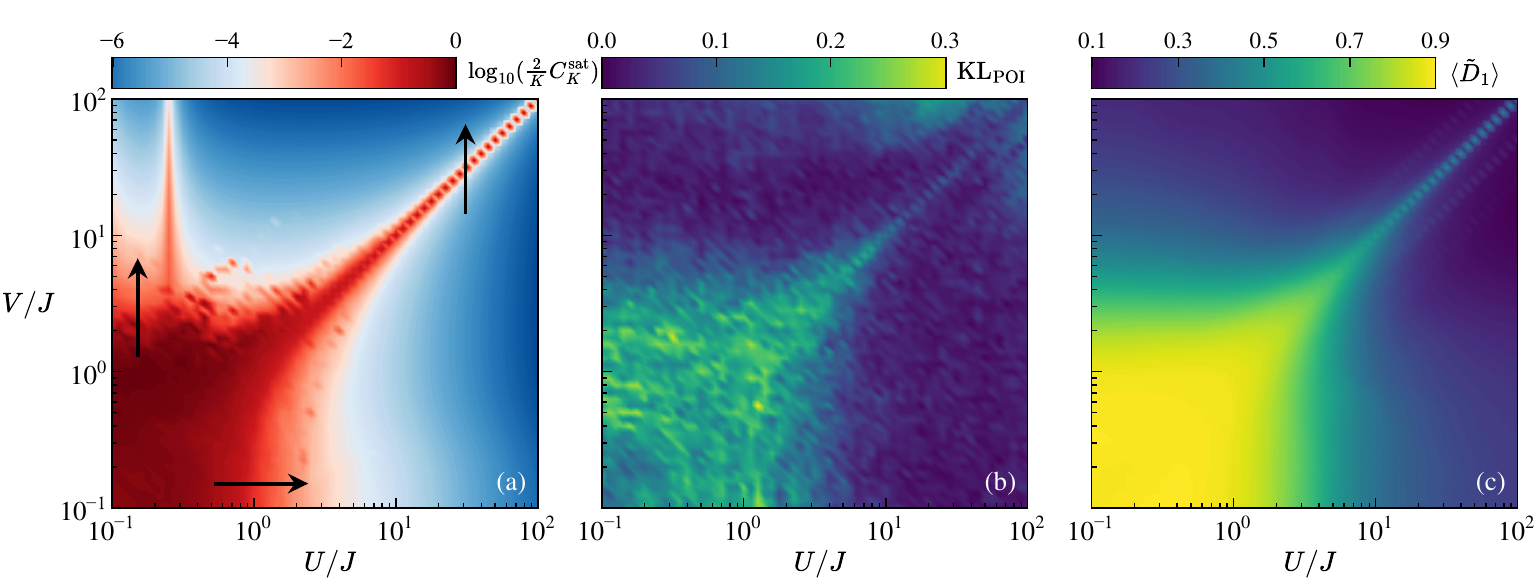}
    \caption{
Complexity diagnostics in the $(U/J,V/J)$ plane for the tilted extended Bose--Hubbard chain. 
(a) Long-time saturation value of the Krylov complexity,
$\log_{10}(2C_K^{\mathrm{sat}}/K)$, for the localized initial state $|\psi_l\rangle$. 
The black arrows indicate crossover regions associated with increasingly intricate Krylov complexity patterns. 
(b) Kullback--Leibler divergence $\mathrm{KL}_{\mathrm{POI}}$. 
(c) Spectrum-averaged participation entropy $\langle\widetilde{D}_1\rangle$. 
 Throughout the calculation, the tilt is $F=0.1J$ and unit filling with $N=L=6$ is used.
}
\label{fig:1}
\end{figure*}

\subsection{Lanczos-sequence and spectral diagnostics}

Beyond the complexity $C_K(t)$, a useful state-sensitive diagnostic is the fluctuation structure of the off-diagonal Lanczos coefficients. We quantify it through the
variance of the logarithmic ratios of the Lanczos coefficients
~\cite{hashimoto2023},
\begin{equation}
\label{eq:sigmalog}
\sigma_{\rm log}^{2}
=
\mathrm{Var}\!\left[
\ln\!\left(\frac{b_{n-1}}{b_n}\right)
\right],
\end{equation}
where $b_n>0$
(boundary terms involving $b_0=0$ are excluded). Large $\sigma_{\rm log}^{-1}$ corresponds to small fluctuations of neighboring logarithmic ratios and hence a smoother Lanczos sequence, while smaller values indicate stronger structure or oscillations. 

For comparison with standard spectral diagnostics, we evaluate the ordered many-body energies $E_n$ and the consecutive spacings $S_n=E_{n+1}-E_n$. We then calculate the consecutive-gap ratio~\cite{Oganesyan2007,Atas2013,hashimoto2023,yanSignaturesQuantumChaos2023},
\begin{equation}
\label{eq:r}
r_n=\frac{\min(S_n,S_{n+1})}{\max(S_n,S_{n+1})},
% \qquad 0\le r_n\le1,
\end{equation}
whose distribution is Poisson (POI)-like for integrable spectra and Gaussian orthogonal ensemble (GOE)-like for chaotic regimes. The corresponding mean values are $\langle r\rangle_{\mathrm{POI}}=2\ln2-1\approx0.3863$ and  $\langle r\rangle_{\mathrm{GOE}}\approx0.5359$~\cite{Atas2013}. 
For integrable systems distribution $P(r)$ follows the Poisson form, whereas for chaotic systems it
approaches the Wigner--Dyson (WD) distribution~\cite{Wigner1993}. 
For POI spectra, we have $P_{\mathrm{POI}}(r)=2/(1+r)^2$. 
% While $P_{\mathrm{GOE}}(r)=27(r+r^2)/[4(1+r+r^2)^{5/2}]$ for GOE  matrices.
The Kullback--Leibler (KL)~\cite{kullback1951information} divergence provides a comparison between the distributions of the model and Poisson prediction, $P_{\mathrm{POI}}(r)$
\begin{align}
\text{KL}_{\mathrm{POI}}=\int_0^1drP\ln\left(\frac{P}{P_{\rm POI}}\right).
\end{align}
It shows how close distribution $P$ to $P_{\rm POI}$. Function $\text{KL}_{\mathrm{POI}}$  becomes zero only when the two distributions are identical. 

The emergence of spectral chaos also entails a fundamental change in the structural properties of the eigenstates, which can be efficiently characterized using finite-size generalized fractal dimensions (GFDs)~\cite{PhysRevLett.122.106603,PhysRevB.84.134209}. We probe Shannon information entropy of the state intensities in Fock space using the finite-size GFDs,
\begin{equation}
\label{eq:d1}
\widetilde{D}_1^{(m)}
=
-\frac{1}{\ln\mathcal{D}}
\sum_{n=0}^{\mathcal{D}-1}
|\psi_n^{(m)}|^2
\ln|\psi_n^{(m)}|^2,
\end{equation}
 % determining the scaling of the Shannon information entropy of the state intensities. 
 where $\psi_n^{(m)}$ is the amplitude of the $m$th eigenstate $|\psi^{(m)}\rangle$ on the number basis state $|n\rangle$. It  measures the localization (delocalization) of the state $|\psi\rangle$ in  basis $|n\rangle$. 
% Here $\psi_m=\langle|\psi_m\rangle$ and $\mathcal{D}$ is the Fock-space dimension. 
We will calculate the spectral average
\begin{equation}
\langle\widetilde{D}_1\rangle
=
\frac{1}{\mathcal{D}}
\sum_{m=1}^{\mathcal{D}}
\widetilde{D}_1^{(m)},
\end{equation}
evaluated over the full eigen spectrum~\cite{Halsey1986}.

\section{Krylov complexity and quantum chaos}\label{sec:results}
\subsection{Krylov complexity vs two-body interactions}

Quantum chaos is typically predicted by the level statistics. However the route to quantum chaos in the tilted extended Bose--Hubbard chain is subtle and requires a careful analysis of the dynamics. Panel~(a) in Fig.~\ref{fig:1} gives the long-time saturation value of the Krylov complexity for initial state $|\psi_l\rangle$, normalized with respect to $\log_{10}(2C_K^{\mathrm{sat}}/K)$, where the normalization factor $K/2$ is the Krylov complexity of the completely delocalized state in the large $K$ limit~\cite{Scialchi2024}. In the weak interaction regime (red region), where both $U/J$ and $V/J$ are small, the dynamics is predominantly governed by tunneling, and the Krylov complexity remains relatively high. When the interaction is very weak,  Krylov dynamics oscillates with large amplitudes and periods where the average instead of saturation is used. For sufficiently strong interactions, however, the Krylov complexity decreases, suggesting that strong interactions suppress transitions of the many-body states and effectively constrain the accessible dynamical subspace. The black arrows indicate crossover regions characterized by increasingly intricate structures in the Krylov complexity landscape. 

%As the interaction strengths increase, the competition between the on-site and  nearest-neighbor interaction  connects strongly  accessible Hilbert-space states, leading to a pronounced increase in the saturation value of the Krylov complexity. 

A prominent diagonal ridge is observed in Fig.~\ref{fig:1}(a) when $U/J$ and $V/J$ are both large and have comparable magnitude.  The comparable competition between the on-site and  nearest-neighbor interaction  connects strongly  accessible Hilbert-space states, leading to a pronounced increase in the saturation value of the Krylov complexity than the neighboring region.  
% This feature indicates an enhanced Krylov complexity arising from the competition between the local and extended interactions. 
Interestingly, this structure is not equally pronounced in the spectral diagnostics shown in Fig.~\ref{fig:1}(b) and~(c). Fig.~\ref{fig:1}(b) displays the KL divergence of the spacing-ratio distribution  $\mathrm{KL}_{\mathrm{POI}}$, while panel~(c) shows the spectrum-averaged participation entropy $\langle \widetilde{D}_1\rangle$. Being consistent with the measuring of averaged spacing ratio~\cite{kollathStatisticalPropertiesSpectrum2010}, these spectral measures exhibit smoother crossover patterns and do not fully reproduce the diagonal enhancement observed in the Krylov complexity map. $\mathrm{KL}_{\mathrm{POI}}$ hardly shows the diagonal ridge in the parameter plane when $U$ and $V$ is large. This discrepancy suggests that Krylov complexity is sensitive to 
initial state, that is not entirely captured by spectral
statistics. The
Krylov complexity thus captures state-dependent information scrambling pathways, which provides complementary information about the dynamical
features of the Hilbert space beyond spectral diagnostics alone.

\subsection{Bose-Hubbard model regime}

\begin{figure}[t]
    \centering
    \includegraphics[width=0.97\linewidth]{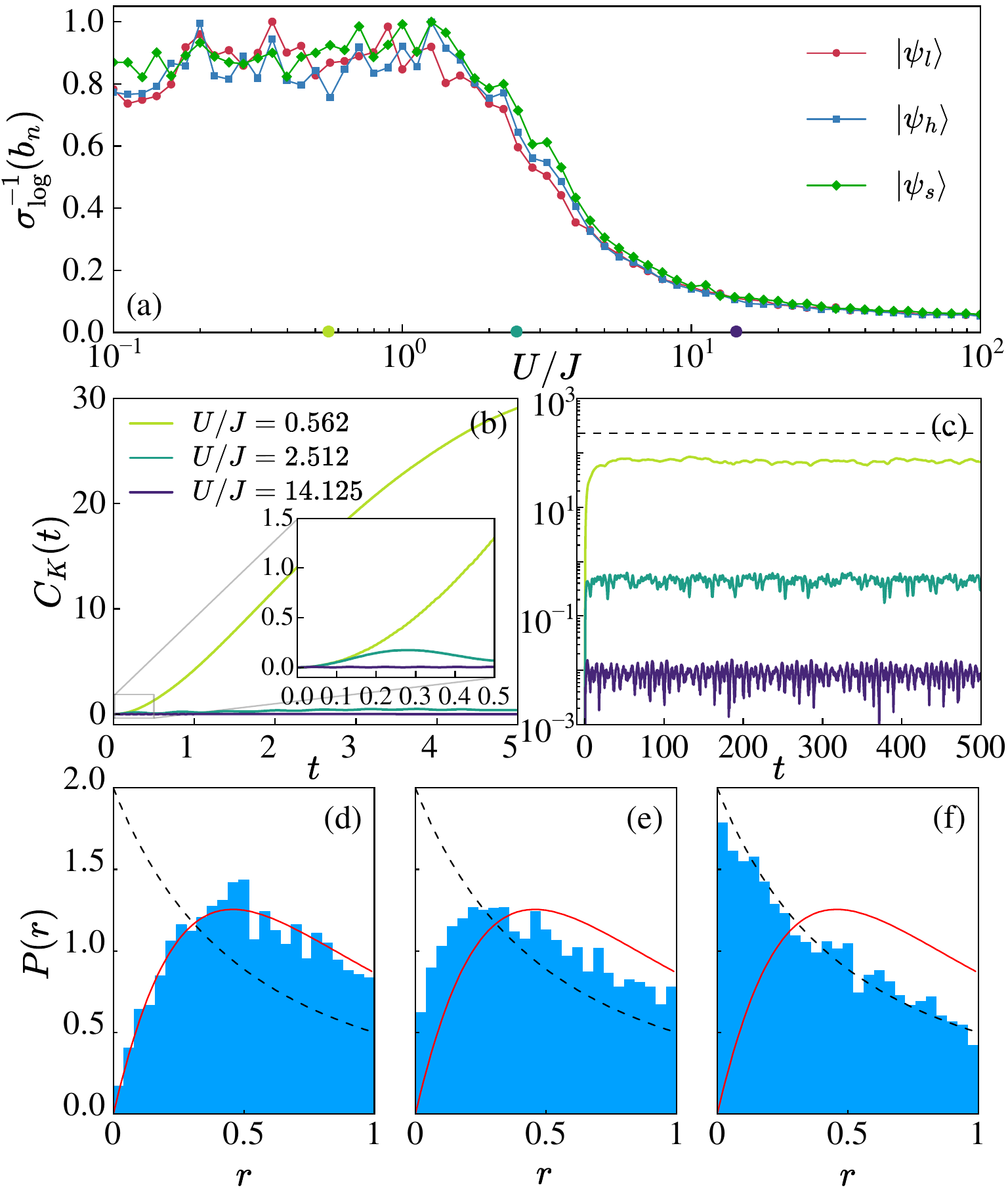}
    \caption{Bose-Hubbard model regime. 
    (a) Lanczos-sequence diagnostic $\sigma_{\log}^{-1}(b_n)$ as function of $U/J$ for the three initial states.  Three dots on the $U/J$ axis correspond to  $U/J=0.562,\, 2.512$ and 14.125, respectively.
    (b) and (c) Time evolutions of the Krylov complexity $C_K(t)$  for $|\psi_l\rangle$ at the three different $U/J$ values. In panel(c), the dashed line gives $K/2$.  Panels
    (d), (e) and (f) show
    level-spacing-ratio histograms at the interaction strengths marked in (a). Dashed black and solid red lines are the Poisson and WD distributions. In all panels $L=6$, $F/J=V/J=0.1$. In the numerical calculations, we consider $N=6$ in (a)--(c) and $N=12$ in (d)--(f). Larger $N$ is used in (d)--(f) to better visualize the distribution of the level statistics.
    }
    \label{fig:2}
\end{figure}

We now turn to the Bose-Hubbard limit with $V\sim0$ to give a reference case. 
Fig.~\ref{fig:2}(a) shows the corresponding normalized inverse logarithmic spread of the Lanczos coefficients, $\sigma_{\log}^{-1}(b_n)$ for three different initial states and relatively weak $V/J=0.1$. The diagnostic is large over the weak and intermediate interaction regimes, where we have a broad distribution of Lanczos coefficients. It decreases $U/J> 1$, where the strong on-site interaction suppresses tunneling of the dressed atoms involving double (and multiple) occupancy. Although the three initial states exhibit a similar trend, their detailed fluctuations and crossover behavior differ. The differences are visibly especially when $U/J<10$, due to the dependence on the initial states.

The time evolution of the Krylov complexity $C_K(t)$, shown in Fig.~\ref{fig:2}~(b) and~(c), exhibits a pronounced dependence on the on-site interaction. At short-times, the complexity grows quadratically with time. However the weak interaction $U/J=0.562$ reaches a much larger Krylov complexity than the intermediate and strong interaction cases ($U/J=2.512,\,14.125$). In Fig.~\ref{fig:2}~(c), the long-time dynamics display persistent fluctuations around a saturated value, reflecting the finite Hilbert-space dimension.
 % and the absence of a clear thermodynamic relaxation regime. 
 For sufficiently strong interactions, the Krylov complexity remains comparatively small, consistent with the suppression of the tunneling processes and the resulting restriction of the accessible many-body states. 

% These results indicate a crossover in the degree of dynamical mixing, but the finite-size data do not support the extraction of a universal Krylov complexity growth law.

% The weak-interaction regime exhibits substantial deviations from the Poisson distribution and shows good agreement with the Wigner--Dyson (WD) distribution, indicating pronounced level repulsion and a nontrivial degree of spectral mixing already in the bare model. As the interaction strength increases, the spectral distribution remains strongly non-Poissonian over the intermediate-interaction regime, while the large-$U/J$ case develops an enhanced weight at small spacing ratios, signaling a gradual suppression of level repulsion and a crossover toward more regular spectral behavior.

Figs.~\ref{fig:2}(d)--(f) are the level-spacing-ratio distributions for $U/J=0.562$,
$2.512$, and $14.125$, respectively. To obtain clearer statistics we consider $N=12$ in the calculation. The dashed black and solid red lines denote
the Poisson and WD reference distributions. In the weak-interaction regime, the distribution deviates
substantially from Poisson and agrees well with the WD form, indicating pronounced level
repulsion and nontrivial spectral mixing
[Fig.~\ref{fig:2}(d)]. The intermediate case [Fig.~\ref{fig:2}(e)] maintains the level repulsion. However it is shifted from the WD to the Poisson distribution. On the other hand, the large-$U/J$ case [Fig.~\ref{fig:2}(f)] retains a
pronounced contribution at small ratios and agrees well with the Poisson distribution. The above result already shows that the state-resolved Krylov
complexity provides complementary information about the regular-to-chaotic
crossover in the finite Bose--Hubbard chain~\cite{Santos2010,Kolovsky2004,Hiller2006,Scialchi2024,Baggioli2025}.

\subsection{Nearest-neighbor interaction dominant regime}

\begin{figure}[t]
    \centering
    \includegraphics[width=0.97\linewidth]{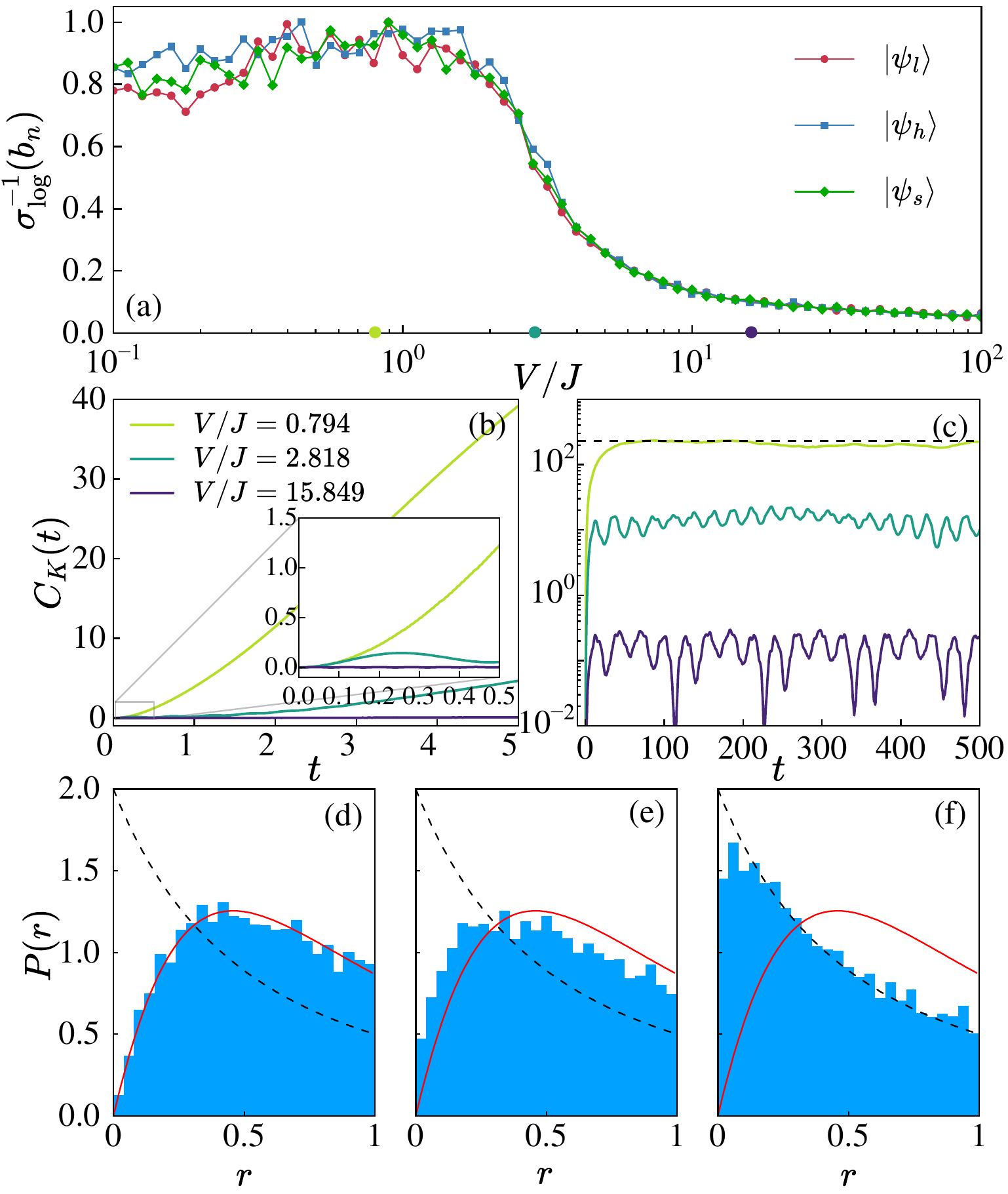}
    \caption{
    Same as Fig.~\ref{fig:2}, but for weak on-site interaction $U/J\simeq 0$.
(a) $\sigma_{\log}^{-1}(b_n)$ as a function of $V/J$ for the three initial  states.  The
dots mark $V/J=0.794$, $2.818$, and $15.849$ on the $V/J$ axis. (b) and (c) Time evolution of $C_K(t)$ for
$\ket{\psi_l}$ at the three interactions marked in (a). When $V/J=0.794$, the corresponding long-time complexity saturates around the dashed line $C_K=K/2$. (d)--(f) Level-spacing-ratio
histograms at the marked interaction strengths. In all panels $L=6$, $F/J=U/J=0.1$, with $N=6$ in (a)--(c) and
$N=12$ in (d)--(f).
    } 
    \label{fig:3}
\end{figure}

We now turn to the investigation of the extended Bose-Hubbard model where the nearest-neighbor interaction is not negligible. We first examine how the dynamics changes when the nearest-neighbor interaction dominates. 
Fig.~\ref{fig:3}(a) shows $\sigma_{\log}^{-1}(b_n)$ as a function of $V/J$ for the three initial states with $U/J=0.1$. 
The diagnostic again develops a broad peak at intermediate $V/J$. The three initial states follow a similar trend, especially when $V/J>1$. Compared to the Bose-Hubbard regime, the data show better agreement when increasing $V/J$. When $V/J<1$ their fluctuations differ, revealing directly the state dependence. 
% The diagnostic remains relatively large over the weak and intermediate interaction regimes, indicating a comparatively broad distribution of Lanczos coefficients. It decreases markedly at sufficiently large $V/J$, where the strong nearest-neighbor interaction suppresses tunneling processes involving double occupancy. Although the three initial states exhibit a similar overall trend, their detailed fluctuations and crossover behavior differ, demonstrating that this smoothness measure is dependent on the choice of Krylov seed.

% The time evolution of the Krylov complexity $C_K(t)$, shown in Fig.~\ref{fig:3}~(b) and~(c), exhibits a pronounced dependence on the nearest-neighbor interaction $V/J$. In short-time dynamics, a weak nearest-neighbor interaction $V/J=0.794$ reaches a much larger Krylov complexity than the other two intermediate and strong nearest-neighbor interactions ($V/J=2.818,\,15.849$), while the remaining trajectories remain confined to relatively small values of $C_K(t)$; see Fig.~\ref{fig:3}~(b). In Fig.~\ref{fig:3}~(c), the long-time dynamics display persistent fluctuations, reflecting the finite Hilbert-space dimension and the absence of a clear thermodynamic relaxation regime. For sufficiently strong interactions, the Krylov complexity remains comparatively small, consistent with the suppression of resonant tunneling processes and the resulting restriction of the accessible Hilbert-space pathways. 

The time evolution of $C_K(t)$, shown in Figs.~\ref{fig:3}(b) and \ref{fig:3}(c),
depends pronouncedly on $V/J$. At short times the weak nearest-neighbor interaction
$V/J=0.794$ reaches a much larger complexity than the intermediate and strong cases
($V/J=2.818,\,15.849$). Compared to the Bose-Hubbard regime, the complexity at $t=5$ is larger, as shown in Fig.~\ref{fig:3}(b). The long-time dynamics
[Fig.~\ref{fig:3}(c)] display similar trend as the Bose-Hubbard model, e.g. the complexity decreases as the interaction $V/J$ increases. However the saturation show sizable difference compared to the Bose-Hubbard case, e.g. they are  larger overall, indicating that the nearest-neighbor interaction enhances the accessible Hilbert-space connectivity. 

%, on one hand.  fluctuations, again reflecting the finite
%Hilbert-space dimension, and the complexity remains comparatively small for sufficiently
%strong interactions. Compared to the bare Bose--Hubbard case, the fluctuations oscillate more strongly and slowly,  and their coupling becomes relatively weak. 

Fig.s~\ref{fig:3}(d)--(f) show the level-spacing-ratio distributions for
$V/J=0.794$, $2.818$, and $15.849$, with $N=12$ and other parameters as in
Fig.~\ref{fig:3}(a). Both in the weak regime [Fig.~\ref{fig:3}(d)] and intermediate case
[Fig.~\ref{fig:3}(e)] the distribution differ substantially from
Poisson and agrees well with the WD distribution.  This is consistent with the dynamics of the complexity, which reaches higher values than the Bose-Hubbard regime.  For the
large-$V/J$ case [Fig.~\ref{fig:3}(f)] we find that the distribution shows good agreement with the Poisson distribution. Overall the nearest-neighbor interaction alone plays largely similar roles as the on-site interaction in the spectral statistics.

\subsection{Comparable on-site and nearest-neighbor interactions}

\begin{figure}[t]
    \centering
    \includegraphics[width=0.97\columnwidth]{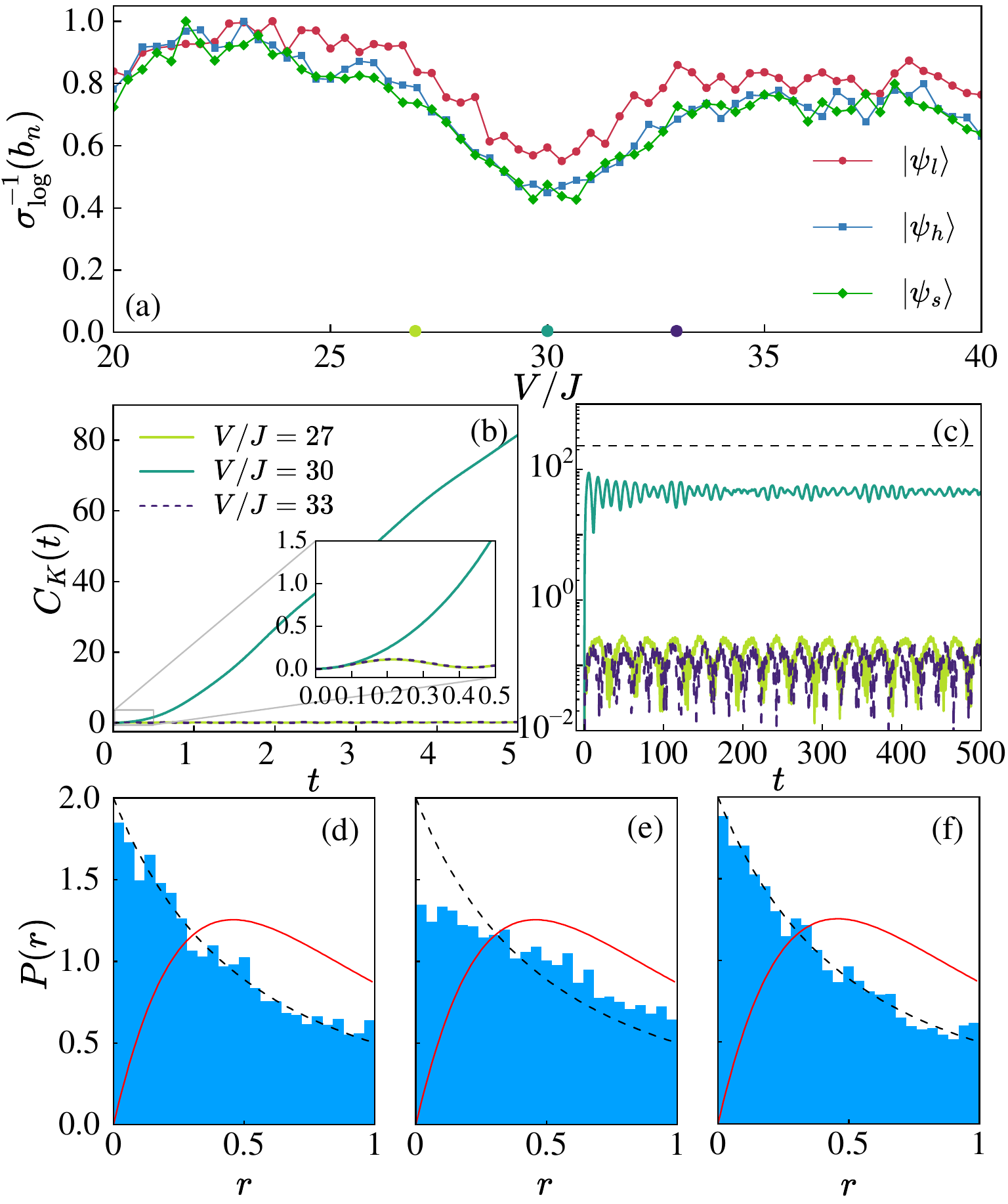}
    \caption{
(a) $\sigma_{\log}^{-1}(b_n)$ as a function of $V/J$ for the three initial Fock states with $U/J=30$; the
dots mark $V/J=27$, $30$, and $33$. (b), (c) Time evolution of $C_K(t)$ for $\ket{\psi_l}$ at
the three interactions marked in (a). The saturated complexity is smaller than the delocalized value $K/2$ marked by the dashed line. (d)--(f) Level-spacing-ratio histograms at the marked
interaction strengths; dashed black and solid red lines are the Poisson and WD distributions.
In all panels $L=6$, $F/J=0.1$, with $N=6$ in (a)--(c) and $N=12$ in (d)--(f).
    }
    \label{fig:4}
\end{figure}

We now investigate the regime when $U$ and $V$ are comparable. As $V$ is increased along the condition $U/J=30$, the Krylov diagnostics identify a pronounced enhancement of many-body spreading. The normalized inverse logarithmic spread is enhanced in the weak to intermediate part of the diagonal scan, and the representative complexity traces show state-dependent spreading followed by saturation and oscillations [Fig.~\ref{fig:4}(a)]. Here the decreasing trend is not found when increasing $V$. The time evolution of $C_K(t)$, shown in Figs.~\ref{fig:4}(b) and \ref{fig:4}(c),
exhibits a pronounced dependence on $V/J$. At short times the comparable interaction case
$V/J=30$ reaches a much larger complexity than the neighboring values (e.g. $V/J=27,\,33$), whose
trajectories remain confined to relatively small values [Fig.~\ref{fig:4}(b)]. 

The long-time dynamics [Fig.~\ref{fig:4}(c)] display persistent fluctuations, reflecting the
finite Hilbert-space dimension. The saturated value is apparently smaller than the largest value (dashed line) that the system can reach. Figs~\ref{fig:4}(d)--(f) are the level-spacing-ratio distributions for $V/J=27$,
$30$, and $33$, with $N=12$ and other parameters as in Fig.~\ref{fig:4}(a).
Fig.~\ref{fig:4}(d) and (f) agree well with the Poisson distribution. The complexity dynamics is relatively low, consistent with the level spacing statistics. This is difference from the variance of the Lanczos coefficients in Fig.~\ref{fig:4}(a), which is high for both parameters. The case in
Fig.~\ref{fig:4}(e) is shifted away from both  distributions. This may  indicate that
 level repulsion and nontrivial spectral mixing become non-negligible. It is consistent with the complexity dynamics shown in Fig.~\ref{fig:4}(c), where the corresponding complexity saturates at a larger value. The variance of the Lanczos coefficient is relatively low. This discrepancy shows that the variance of the Lanczos coefficients can not accurately predict the dynamical properties in this regime.

\section{Discussion}
\label{sec:discussion}
\begin{figure}[t]
    \centering
    \includegraphics[width=0.97\linewidth]{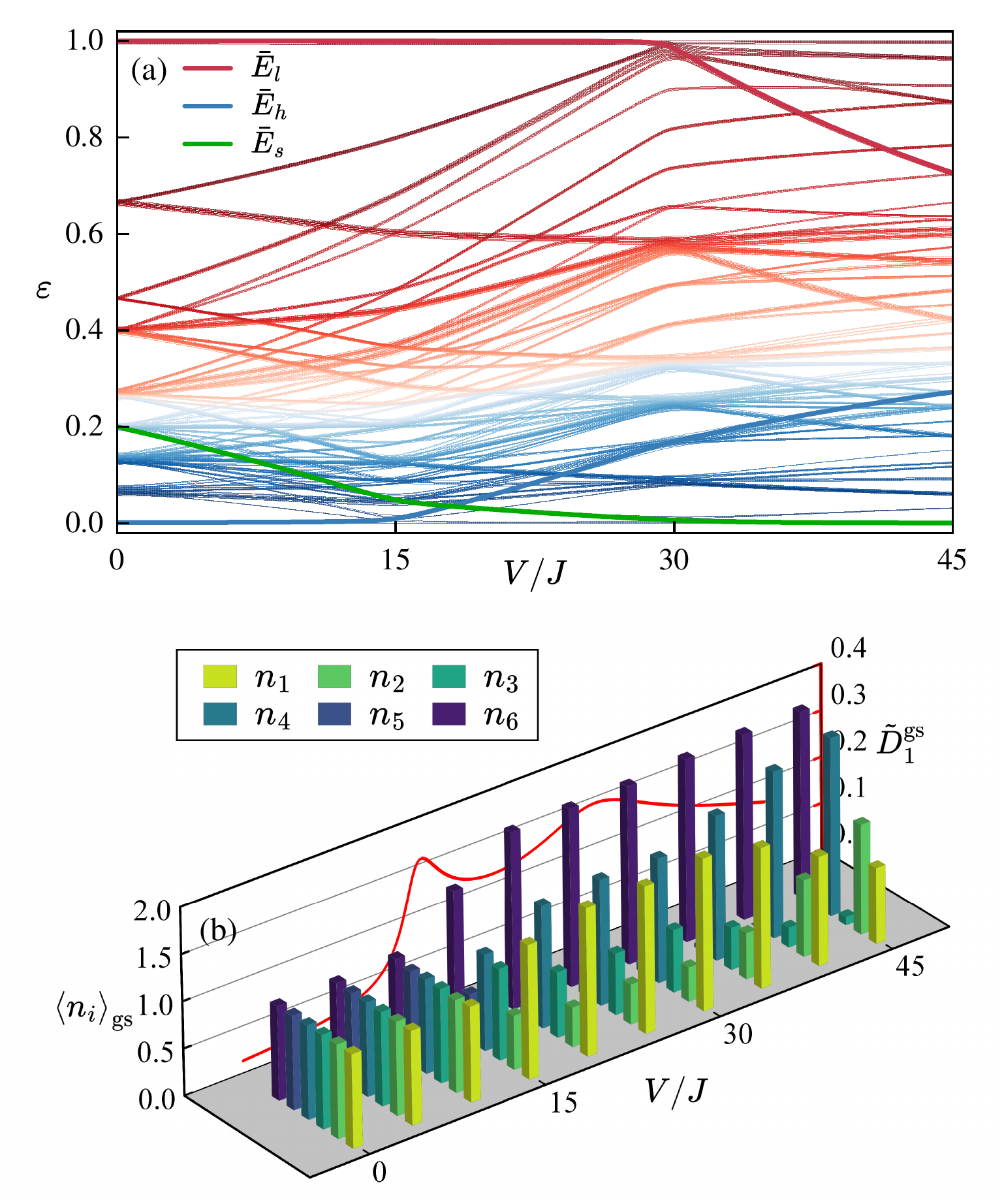}
    \caption{
    (a) Scaled energy as a function of $V/J$ for the three initial states with
$U/J=30$. (b) Ground-state occupation of each site as a function of $V/J$.  The red line in
the side panel denotes the ground-state GFD. It becomes large when $V\sim U$. Other parameters are $F/J=0.1$ and
$L=N=6$.
    }
    \label{fig:5}
\end{figure}

%The strong quantum chaotic behavior around $V\sim U$ (i.e. diagonal ridge) is particularly pronounced in the Krylov complexity map of Fig.~\ref{fig:1}(a). Such regime has multiple consequences for the dynamics as well as ground state of the system. With the given initial states, strong level mixing takes place. 

 When $U$ and $V$ are simultaneously large and comparable, neither local clustering nor period-two density ordering is overwhelmingly favored.  On-site repulsion suppresses multiple occupancy, while nearest-neighbor repulsion penalizes adjacent occupancy. The  competition between these two ordering tendencies frustrates and generates a dense manifold of near-resonant many-body configurations. To demonstrate this, we numerically analyze scaled eigenenergy~\cite{Clavero2025,PhysRevLett.126.150601},
\begin{equation}
\varepsilon=\frac{E-E_{\rm min}}{E_{\rm max}-E_{\rm min}},
\end{equation}
where $E_{\rm max}$ and $E_{\rm min}$ denote the maximum and minimum eigenenergies for a given set of system parameters. 
We also show the initial state trajectories in the scaled energy space~\cite{SciPostPhys.5.5.045,10.1145/1089014.1089019}, as determined by ${\bar E}_j=\langle\psi_j|\hat{H}|\psi_j\rangle$ ($j=l,\,h,\,s$)~\cite{Pausch_2025}.
We fix $U/J=30$ and examine the dependence on $V/J$. Fig.~\ref{fig:5}(a) 
shows the scaled energies together with the
initial-state trajectories ($\bar E_l$, $\bar E_h$, $\bar E_s$) associated with
$\ket{\psi_l}$, $\ket{\psi_h}$, and $\ket{\psi_s}$, embedded in the surrounding many-body
levels. The three states respond qualitatively differently to the increasing nearest-neighbor
interaction: $\bar E_l$ remains nearly unchanged for $V<U$ and varies pronouncedly around
$V\simeq U$, whereas $\bar E_h$ and $\bar E_s$ change strongly already around $V\simeq U/2$. Interestingly, they all show level crossings (avoided crossings) with other eigenstates around $U\approx V\approx 30$. 
The complex level crossings around $U\sim V$ are accompanied by the enhanced Krylov complexity observed in Fig.~\ref{fig:1}(a). However they tend to cross with a small faction of the eigenenergies, preventing the dynamics to be fully chaotic. 

Finally we note that this parameter regime also alters the ground-state properties. When $V\ll U$ and $J\ll U$, the ground-state is Mott insulator phase $|\psi_{\rm gs}\rangle\approx |111111\rangle$~\cite{Fisher1989,Jaksch1998,Greiner2002}. Each site is occupied by a single boson, while the tilt hardly changes the uniform density, as the creation of doublon-hole pair costs energy $\sim U\gg FL$. When $V$ becomes comparable to $U$, the nearest-neighbor interaction starts to play a significant role, and alters the ground-state phases from a Mott insulator to a density-wave phase~\cite{Lahaye2009}. To show this, we estimate the interaction energy of state $|111111\rangle$ is $5V$, while the interaction energy of period-2 charge-density wave states $|202020\rangle$ and  $|020202\rangle$ is $3U$. Balancing the two interaction energies, we obtain $V\approx 3U/5=18$. As shown in Fig.~\ref{fig:5}(b), the site-resolved ground-state
occupations $\langle n_i\rangle_{\mathrm{gs}}$ vary from uniform to the density-wave pattern when $V/J\sim 18$. Such non-uniform distribution becomes even significant when $V\approx U=30$. Further increasing $V/J$, the density profile becomes asymmetric, as the tilt breaks the symmetry. In the thermodynamical limit, the ground-state may undergo a phase transition when increasing $V/J$. Whether one can identify notable signatures of this transition in the dynamics (e.g. Loschmidt echo~\cite{heylScalingUniversalityDynamical2015}) by quenching $V/J$ across $V\sim U$ is an interesting question and worth further investigation. 

\section{Conclusion}
\label{sec:conclusion}

We have investigated Krylov state complexity in a tilted and finite extended Bose-Hubbard chain.
% ~\cite{weckesserRealizationRydbergdressedExtended2025}.
Exact diagonalization and Lanczos recursion from three Fock states show that the time-dependent complexity, its long-time average, and the normalized inverse Lanczos fluctuation provide information complementary to consecutive-gap ratios and Fock-basis participation entropy. For the Bose-Hubbard model, the chaotic regime emerges at intermediate on-site interaction, consistent with earlier spectral studies~\cite{Kolovsky2004,Hiller2006,Santos2010} and recent Krylov operator complexity calculations~\cite{Bhattacharyya2023}. Taking into account of the nearest-neighbor repulsion,  the chaotic window becomes broad, and Krylov-space spreading is enhanced. An important finding is the diagonal ridge near $U\sim V$, where Krylov state spreading is dramatically enhanced. We have argued that this arises because competing on-site and nearest-neighbor interaction generates a dense manifold of near-resonant configurations with other eigenstates. This regime demonstrates unique features of state complexity over spectral diagnostics, as it captures state-dependent information scrambling and can identify strong dynamical mixing. 

%Our results are consistent with recent work establishing Krylov complexity as an order parameter for quantum chaotic transitions~\cite{Baggioli2025}. They also complement operator-growth studies in small Bose-Hubbard clusters~\cite{Bhattacharyya2023} by revealing how competing interaction ranges generate novel dynamics in extended Bose-Hubbard chains.

Our study indicates that Krylov state complexity and spectral chaos diagnostics can
diverge in finite interacting lattice systems~\cite{Santos2010,DAlessio2016}, and complement
work dissecting the relationship between Krylov complexity growth and information
scrambling in many-body systems~\cite{Chen2025}. It is interesting to explore how the complexity is affected by the long-range hopping~\cite{PhysRevLett.131.203003} in general Bose-Hubbard model through the dipolar couplings~\cite{leseleucObservationSymmetryprotectedTopological2019,zhangObservationNonHermitianManybody2025a}. The sensitivity of the Krylov state complexity to initial conditions makes it valuable for understanding how specific experimental preparations explore many-body Hilbert space, complementing ensemble-averaged spectral statistics and opening new avenues for characterizing the onset of quantum chaos in ultracold neutral atom platforms.

\textit{Acknowledgements.} W. L. acknowledges financial support from the EPSRC (Grant No. EP/W015641/1), UKRI UKRI3702 (QCS) and the University of Nottingham. Y. C. acknowledges supports from Taishan College, Shandong University.  L. Q. acknowledges the National Natural Science Foundation of China (NSFC) under Grants No. 12247146, and the Program of China Scholarship Council under Grants No. 202508410217.

%%% SECTION
%\section{Data availability.}

\textit{Data availability.} The data that support the findings of this work are openly available \cite{code}.

%%% SECTION
%\section{AI usage statement.}

\textit{AI usage statement.} The authors used \emph{Anthropic}'s \emph{Claude (Sonnet 5)} for grammar and spell-checking of the manuscript. All suggestions were reviewed and approved by the authors.

%The results connect to the broader understanding of how Krylov-based measures of state spreading relate to thermalization: Krylov complexity is sensitive to the choice of initial state and
%ensemble-averaged Krylov diagnostics do not by themselves generically characterize
%chaos~\cite{SreeramPG2025}, while thermalization in the Krylov basis has been studied as a
%framework linking state spreading to steady-state behavior~\cite{Alishahiha2025}. 

% The extended Bose-Hubbard chain thus provides a clean example of how state-dependent Krylov complexity and global spectral chaos diagnostics can diverge in finite interacting lattice systems~\cite{Santos2010,DAlessio2016}. 

%This dissociation has important implications for interpreting thermalization in quantum simulators: a system may exhibit strong information scrambling from experimentally relevant initial states without satisfying strict ETH criteria across its full spectrum. Because our
%analysis is at $L=6$, the quasi-chaotic diagonal regime should be viewed as a finite-size
%signature, but its mechanism---non-trivial competition between interaction types---is robust
%and likely to persist at larger sizes, possibly with modified coupling thresholds.

%From a broader perspective, our work demonstrates that Krylov state complexity is a powerful and experimentally relevant diagnostic for quantum chaos, thermalization, and information scrambling in realistic quantum simulators. 

%\bibliographystyle{apsrev4-2}

%

\end{document}